\documentclass{siamonline1116}

\usepackage{lipsum}
\usepackage{amsfonts}
\usepackage{graphicx}
\usepackage{epstopdf}
\usepackage{algorithmic}
\ifpdf
  \DeclareGraphicsExtensions{.eps,.pdf,.png,.jpg}
\else
  \DeclareGraphicsExtensions{.eps}
\fi

\numberwithin{theorem}{section}

\newcommand{\TheTitle}{REM sleep complicates period adding bifurcations from monophasic to polyphasic sleep behavior in a sleep-wake regulatory network model for human sleep} 
\newcommand{\ShortTitle}{REM sleep complicates the bifurcations from monophasic to polyphasic sleep}
\newcommand{\TheAuthors}{K. Kalmbach, V. Booth, and C. G. Diniz Behn}

\headers{\ShortTitle}{\TheAuthors}

\title{{\TheTitle}\thanks{Submitted to the editors DATE.
\funding{This work was funded by NSF DMS 1412571 (CGDB) and DMS 1412119 (VB).}}}

\author{
  Kelsey Kalmbach \thanks{Department of Applied Mathematics and Statistics, Colorado School of Mines, Golden CO 80401.}
  \and
  Victoria Booth\thanks{Departments of Mathematics and Anesthesiology, University of Michigan, Ann Arbor, MI 48109.} 
  \and
  Cecilia G. Diniz Behn\thanks{Department of Applied Mathematics and Statistics, Colorado School of Mines, Golden CO 80401 (cdinizbe@mines.edu)}.
}

\usepackage{amsopn}

\usepackage{lipsum}
\usepackage{amsfonts}
\usepackage{graphicx}
\usepackage{epstopdf}
\usepackage{algorithmic}
\ifpdf
  \DeclareGraphicsExtensions{.eps,.pdf,.png,.jpg}
\else
  \DeclareGraphicsExtensions{.eps}
\fi

\pdfoutput=1

\ifpdf
\hypersetup{
  pdftitle={\TheTitle},
  pdfauthor={\TheAuthors}
}
\fi

\begin{document}

\maketitle

\begin{abstract}
The structure of human sleep changes across development as it consolidates from the polyphasic sleep of infants to the single nighttime sleep period typical in adults.  Across this same developmental period, time scales of the homeostatic sleep drive, the physiological drive to sleep that increases with time spent awake, also change and presumably govern the transition from polyphasic to monophasic sleep behavior.  Using a physiologically-based,  sleep-wake regulatory network model for human sleep, we investigated the dynamics of wake, rapid eye movement (REM) sleep, and non-REM (NREM) sleep during this transition by varying the homeostatic sleep drive time constants.  Previously, we introduced an algorithm for constructing a one-dimensional circle map that represents the dynamics of the full sleep-wake network model.  By tracking bifurcations in the piecewise continuous circle map as the homeostatic sleep drive time constants are varied, we establish evidence for a border collision bifurcation that results in period-adding-like behavior in the number of sleep cycles per day.  Interestingly, this bifurcation is preceded by bifurcations in the number of REM bouts per sleep cycle that exhibit truncated period-adding-like behavior.  The interaction of these bifurcations in numbers of sleep episodes and numbers of REM bouts per sleep episode generate non-monotonic variation in sleep cycle patterns as well as quasi-periodic patterns during the transition from polyphasic to monophasic sleep behavior.  This analysis may have implications for understanding changes in sleep in early childhood when preschoolers transition from napping to non-napping behavior, and the wide interindividual variation observed during this transition.  
\end{abstract}

\begin{keywords}
  REM sleep, circle map, piecewise smooth dynamical systems, border collision bifurcation
\end{keywords}

\begin{AMS}
  34C15, 34C55, 37N25, 70K70, 92C20
\end{AMS}

\section{Introduction}

Networks of brainstem and hypothalamic neurons produce distinct \\states of wake, rapid eye movement (REM) sleep, and non-REM (NREM) sleep and govern transitions between these states \cite{Saper2001, Saper2005}.  These networks are largely conserved across mammalian species, but differences in network interactions give rise to a great diversity of sleep-wake behavior.  Even within an individual human, the sleep-wake regulatory network produces a range of behavior across the lifespan with rapid cycling between sleep and wake states in infants, regular napping behavior in early childhood, and a consolidated nighttime sleep period in adults \cite{CarskadonDementChap2}.

Both circadian ($\sim 24$ h) and homeostatic sleep drives interact with sleep-wake networks to produce typical sleep-wake behavior.  The circadian regulation of sleep is driven by the circadian pacemaker in the suprachiasmatic nucleus (SCN) \cite{Mistlberger2005, Saper2005}.  In SCN neurons, intrinsic molecular clocks with periods of approximately 24 h are entrained to the 24 h light-dark cycle and produce 24 h oscillations in firing rate.  Projections from the SCN to neuronal populations in the sleep-wake regulatory network affect the timing of sleep and wakefulness and drive nocturnal, diurnal, crepuscular, or other patterns of behavior \cite{Saper2005}.  The homeostatic sleep drive depends on the past history of sleep and wakefulness and reflects sleep need that increases with prolonged wakefulness.  However, observed sleep-wake behavior reflects interactions among circadian and homeostatic drives and sleep-wake network dynamics.  For example, when shift work requires sleep to occur at atypical times of day, circadian and homeostatic interactions typically fail to produce consolidated periods of sleep comparable to those that occur on a standard day-night schedule. 

In recent years, physiologically-based mathematical models have been used to evaluate conceptual models of sleep-wake regulation \cite{BandDB2014, DBandB_SIADS2012, Fleshner, Kumar, PR, Rempe, Tamakawa}.  Models can provide insight into the physiological mechanisms associated with observed differences in sleep-wake behavior and improve understanding of the transitions that occur in sleep-wake dynamics.  Variations in sleep-wake behavior across development and across species provide important constraints for these mathematical models and facilitate broad comparisons of different network architectures \cite{DBAandB_SIADS2013}.  One such developmentally-mediated change occurs in early childhood when children transition from napping to non-napping behavior.  Most two year old children nap regularly, and by five years old,  children typically consolidate their sleep to a single nighttime sleep period that is similar to the adult pattern \cite{Acebo2005, Iglowstein, Jenni2006}.  However, this transition from biphasic to monophasic sleep is marked by large interindividual differences that may reflect both physiological and environmental effects.

Slow wave activity (SWA) in the electroencephalogram (EEG) is an established marker of the homeostatic sleep drive \cite{Rusterholz}, and sleep deprivation experiments have been used to estimate the time constants associated with the homeostatic sleep drive in adults with typical sleep-wake behavior as well as under other conditions.  Experiments have identified differences in the rates of growth and decay of homeostatic sleep drive in different species \cite{Phillips2013, Tobler} and in humans at different life stages \cite{Gaudreau2001, Jenni2005}.  As a first step in understanding the effects of the time scales of the homeostatic sleep drive in our sleep-wake network model, we analyzed model behavior as homeostatic time constants were varied.

Previous work has considered the effects of varying homeostatic time constants in other mathematical models of sleep-wake regulation and examined implications for inter-species differences \cite{Phillips2013} and changes across development \cite{Skeldon}.  These results support a key role for homeostatic time constants in producing distinct patterns of sleep-wake behavior.  In addition, this work also identified period adding structure in the bifurcations produced as homeostatic time constants changed \cite{Skeldon}. 
 However, these models simulate only two behavioral states, wake and sleep, and do not account for alternation between NREM and REM sleep during the sleep period.  It is unknown how NREM-REM dynamics may affect the overall dynamics of the network as the homeostatic time constants change.  

In this paper, we present a computationally-based analysis of the bifurcations produced by varying homeostatic time constants in a sleep-wake network model that simulates three states: wake, NREM sleep, and REM sleep.  Previously, we introduced an algorithm for constructing a one-dimensional circle map that represents the dynamics of the full sleep-wake network model.  By tracking bifurcations in the piecewise continuous circle map, we establish evidence for a border collision bifurcation that results in period adding-like behavior in the average number of sleep cycles per day.  Interestingly, this bifurcation is preceded by bifurcations in the number of REM bouts per sleep cycle that also exhibit period adding-like behavior.  The interaction of these bifurcations in numbers of sleep episodes and numbers of REM bouts per sleep episode generate non-monotonic variation in sleep cycle patterns during the transition from polyphasic to monophasic sleep behavior. 

The paper is organized as follows:  in Section 2 we briefly review the sleep-wake regulatory network model and describe the numerically-constructed one-dimensional circle map that captures some of the key dynamics of the full model; in Section 3 we describe the bifurcations produced by varying the time constants of the homeostatic sleep drive with a specific focus on the transition from monophasic to biphasic sleep; in Section 4 we describe the role of bifurcations in the number of REM bouts within the greater bifurcation structure; and in Section 5 we provide a brief summary of our results and relate them to previous results in two-state models of sleep-wake regulation.


\section{One-dimensional map to describe sleep-wake dynamics}
\label{1Dmap}

In previous work, we developed a physiologically-based sleep-wake network model for human sleep \cite{GDBandB}.  Using fast-slow analysis techniques, we analyzed the dynamics of this model in different parameter regimes \cite{ BXandDB, DBandB_SIADS2012} and showed that the interactions between the slower circadian and homeostatic time scales play a strong role in determining overall model dynamics.  To further investigate circadian modulation of sleep-wake dynamics, we numerically constructed a one-dimensional map that captures salient dynamics of the full model \cite{BXandDB}.  In this section, we present the sleep-wake network model and its associated one-dimensional map.

\subsection{Reduced sleep-wake model}
\begin{figure}[h!]
\centering
\includegraphics[width=\textwidth]{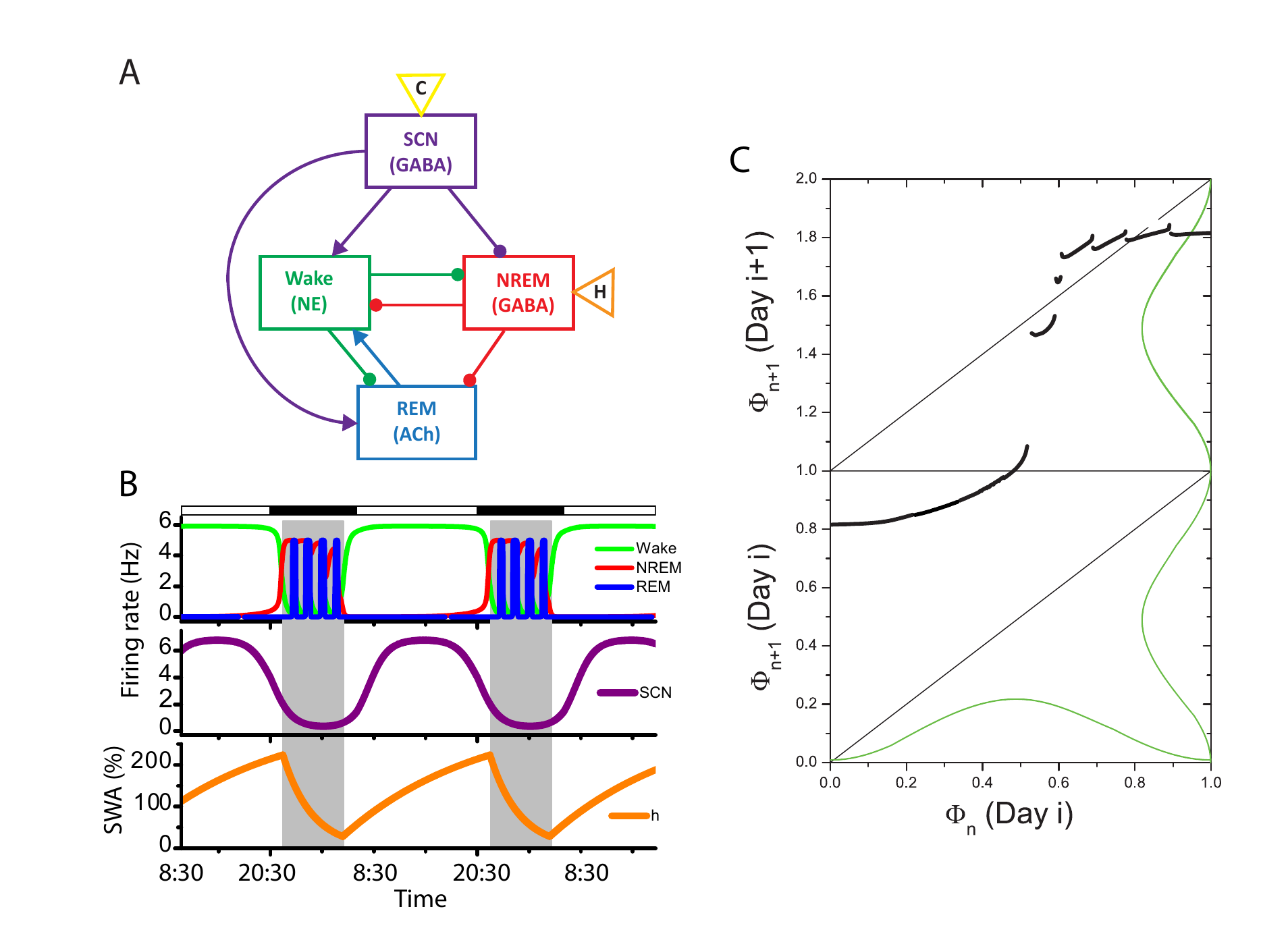}
\caption{Physiologically-based model for sleep-wake regulation.  A.  Schematic of model connectivity summarizes interactions among wake-, NREM sleep-, and REM sleep-promoting populations with circles denoting inhibition and arrows denoting excitation.  B. Under baseline conditions, model simulates adult human sleep with firing rates for wake- (green), NREM sleep- (red), and REM sleep- (blue) promoting populations driving the production of behavioral state.  Above the top panel, black and white bars indicate the timing of light and dark, and sleep episodes are represented by gray shading.  C.  The one-dimensional map captures the dynamics of circadian modulation of the sleep-wake regulatory network model.  The map describes circadian phase of the (n+1)th sleep onset $\Phi_{n+1}$, computed relative to the minimum of the circadian drive variable $c$, as a function of circadian phase of the $n$th sleep onset $\Phi_n$.  The double plot indicates the day on which $\Phi_{n+1}$ occurs, and traces of the circadian drive variable $c$ (green curves) are show for reference.}
\label{fig1}
\end{figure}

Based on leading conceptual models for the structure of the sleep-wake regulatory network, our mathematical model describes neurotransmitter-mediated interactions among neuronal populations that promote the three primary behavioral states of wake, NREM sleep and REM sleep (Figure \ref{fig1}A).  The bidirectional, inhibitory projections between the wake-promoting and NREM sleep-promoting populations reflect the mutually inhibitory flip-flop switch model for sleep-wake transitions \cite{Saper2001} and the excitatory-inhibitory projections between the wake- and REM sleep-populations reflect the reciprocal interaction hypothesis for REM sleep \cite{McCarleyHobson,McCarleyMassaquoi}.  Representative wake-promoting monoaminergic populations include the locus coeruleus (LC) and the dorsal raphe (DR); NREM sleep-promoting populations include GABAergic, sleep-active neurons of the ventral lateral preoptic nucleus (VLPO); and REM-promoting populations include the REM-active, cholinergic areas of the laterodorsal tegmental nucleus (LDT) and pedunculopontine tegmental nucleus (PPT). 

In the reduced population firing rate model formalism, we assume that the mean firing rate of a presynaptic population, $F_Y(t)$, induces instantaneous expression of  neurotransmitter concentration, $C_i(t)$ \cite{DBandB_SIADS2012}.  Thus, $C_i(t)=C_{i\infty}(F_{Yi})$ where the steady state neurotransmitter release function is normalized between 0 and 1 and given by $C_{i\infty}(f)=\tanh(f/{\gamma_i})$ with the parameter $\gamma_i$ controlling the sensitivity of release as a function of presynaptic firing rate $f$. 

Thus, the mean post-synaptic firing rate $F_X(t)$ (in Hz) is governed by the equation
\begin{equation}
F_X'=\frac{F_{X\infty}(\sum_i g_{i,X}C_{i\infty}(F_{Yi}))-F_X}{\tau_X}
\label{FX}
\end{equation}
where the steady state firing rate response function has the sigmoidal profile utilized in standard firing rate models \cite{WilsonCowan} (see reviews in \cite{DayanAbbott,Deco2008,Ermentrout1998}):
\begin{equation}
F_{X\infty}(z)=\frac{X_{max}}{2}(1+\tanh((z-\beta_X)/\alpha_X)).
\label{FXinfty}
\end{equation}
The parameters $X_{max}$, $\alpha_X$, and $\beta_X$ represent the maximum firing rate, sensitivity of response, and half-activation threshold, respectively.  The argument  of $F_{X\infty}(\cdot)$ consists of the sum of effective synaptic inputs generated by the neurotransmitter concentrations released to population $X$ for $X = W$ (wake-promoting), $N$ (NREM sleep-promoting) or $R$ (REM sleep-promoting).  The variable $F_X(t)$ evolves, with time constant $\tau_X$ in minutes, to the value determined by its steady state firing rate response function $F_{X\infty}(\cdot)$.  

In our model \cite{BXandDB}, the circadian rhythm promotes waking and inhibits NREM sleep, consistent with evidence for its role in human sleep behavior \cite{Abrahamson2001, Edgar1993}.  The firing rate of the population describing the SCN,  $F_{SCN}(t)$, is governed by Eq.\ \ref{FX} with $X=SCN$, and its associated neurotransmitter ($i = S$) simulates the effect of indirect synaptic projections from the SCN to the wake-, NREM-, and REM sleep-promoting populations. The argument of the steady state firing rate response function $F_{SCN\infty}(\cdot)$ is the circadian drive $c(t)$ modeled with the circadian oscillator model described by Forger and colleagues \cite{Forger1999,KronauerErrata,SerkhForger}.  This model, based on a modified van der Pol oscillator, includes a circadian drive variable $c(t)$ and a complementary variable $x_c(t)$ governed by the following equations:

\begin{eqnarray}
c' &=& \left(\frac{\pi}{12}\right) (x_c + B) \label{circeq1} \\
x_c' &=& \left(\frac{\pi}{12}\right) \left[\mu \left( x_c - \frac{4x_c^3}{3} \right) -c \left[ \left(\frac{24}{0.99669 \tau_x} \right)^2 + kB \right] \right].
\label{circeq2}
\end{eqnarray}
\noindent

The term $B$ represents light input to the model \cite{Forger1999} and involves a variable $n$ that describes the dynamics of light-induced activation and is governed by $n' = 60[\alpha(I(t))(1-n) - \beta n]$ where $\alpha(I(t)) = \alpha_0 (I(t) / I_0)^p$ and $I(t)$ is light intensity.  The circadian drive variable $c(t)$ oscillates between $\sim -1$ and $\sim +1$ with an intrinsic period of 24.2 h and generates rhythmic increases and decreases in $F_{SCN}$ between $\sim 1$ to $7$ Hz, consistent with experimental measurements of SCN neuronal activity in mammals \cite{Meijer}.  As in our previous work \cite{BXandDB}, we assume the circadian oscillator is entrained to a 24 h environmental light schedule which is simulated by a light input of 500 lux to the circadian drive model on a 14:10 Light:Dark cycle.

The homeostatic sleep drive $h(t)$ describes the exponential growth and decay of sleep propensity as a function of prior sleep-wake history:
\begin{equation}
h'=\frac{(H_{max}-h)}{\tau_{hw}}H[F_W-\theta_W]-\frac{h}{\tau_{hs}}H[\theta_W-F_W]
\end{equation}
where the units are \% mean SWA; $H[z]$ is the Heaviside function defined as $H[z]=0$ if $z<0$ and $H[z]=1$ if $z\ge0$; and the state dependence of homeostatic activity is gated by the firing rate of the $W$ population, $F_W$.  The mechanism of action of $h(t)$ mimics the action of adenosine \cite{Basheer,Huang} and affects excitability of the NREM sleep-promoting population by modulating the half-activation parameter of its steady state activation function according to $\beta_N(h) = -0.0045 h - 0.1$.

The time constants of $h(t)$ were specified to be $\tau_{hw}=15.78$ h and $\tau_{hs}=3.37$ h based on fits to the time course of changes in SWA in adults \cite{Rusterholz}.  For these homeostatic time constants, the model produces approximately 8 hours of sleep and 16 hours of wake each day (Fig. \ref{fig1}B). Simulated sleep is initiated with NREM sleep, and then the model cycles between NREM sleep and REM sleep, resulting in four REM bouts across the sleep period.  The timing of sleep and wakefulness are entrained to the 14:10 Light:Dark cycle \cite{BXandDB}.  Sleep onset occurs on the descending interval of the circadian drive at a circadian phase of $0.793$ (where phase is normalized between zero and one with zero phase corresponding the circadian minimum).  Wake onset occurs just past the circadian minimum at the beginning of the rising interval of the circadian drive at a phase of $0.111$.  The duration, timing, and architecture of the simulated sleep-wake behavior are consistent with normal adult human sleep \cite{CarskadonDementChap2, CzeislerBuxtonChap35, Feinberg}.

\subsection{One-dimensional map for sleep-wake model}

In previous work, we developed a numerical algorithm for computing a one-dimensional map to describe the relationship between the circadian phases of successive sleep onsets \cite{BXandDB}.  We define sleep onset to be the time at which $F_W$, the firing rate of the wake-promoting population, decreases through 4 Hz, and the circadian phase of sleep onset is the time difference between sleep onset and the preceding minimum of the circadian drive variable $c$, divided by the period of the circadian oscillator entrained to the light/dark cycle (24 h).  The map is constructed by tracking circadian phases of sleep onsets during numerical integrations of the model from a large number of initial conditions. 

In order to specify appropriate initial conditions, we consider the model as two subsystems, the sleep-wake network and the circadian oscillator, that are coupled together.  The sleep-wake network includes the neuronal population firing rates ($F_W$, $F_N$ and $F_R$) and the homeostatic sleep drive $h$, while the circadian oscillator includes the circadian variables $c$, $x_c$, and $n$ as well as the SCN firing rate ($F_{SCN}$) since its activity is directly driven by the circadian drive $c$.  Initial conditions are selected to place the sleep-wake variables on a stable solution of the sleep-wake network near sleep onset associated with a fixed value of the circadian drive $c$ and the circadian oscillator variables on their stable solution associated with the same fixed value of $c$.  To construct a full set of initial conditions, we varied $c$ over one circadian cycle (approximately $-1.114 < c < 1.114$) and used two-parameter numerical continuation (implemented in AUTO on XPPAUT \cite{xpp}) to compute the stable solution of the sleep-wake network.  The map is generated by plotting the phase of sleep onset at cycle $n+1$, $\Phi_{n+1}$, as a function of the phase of sleep onset at cycle $n$, $\Phi_n$ (Figure \ref{fig1}C).  All map computations were performed in MATLAB (MathWorks Inc., Natick, MA). For additional details regarding the algorithm for constructing the one-dimensional map, see \cite{BXandDB}. 

The resulting one dimensional map for the sleep-wake model is piecewise continuous and non-monotonic with vertical discontinuities occurring at cusps in the map curves demarcating distinct branches of the map.  The branches of the map correspond to sleep-wake cycles with distinct characteristics including differences in sleep and wake bout durations and number of REM bouts.   The map has a fixed point, seen as an intersection point of the map with the $y = x$ line, near phase $\Phi^s =0.793$. 
The slope of the map at this intersection point is less than 1 (estimated as $0.02$) indicating a stable fixed point \cite{KaplanGlass}.  This fixed point corresponds to the periodic solution associated with stable entrainment of typical adult sleep-wake behavior to the circadian rhythm (Figure \ref{fig1}B).  The evolution of sleep onset phases over multiple cycles from a desynchronized state (where $\Phi_0$ is chosen to be away from this fixed point) to the entrained state corresponding to the fixed point can be traced out on the map by the usual ``cobwebbing" technique.  Thus, we can predict cycle-by-cycle changes in sleep-wake behavior during the re-entrainment process by tracking the model trajectory on the map.

\section{Bifurcations in the map as dynamics of the homeostatic sleep drive vary}
\label{3StateBifns}

In this study, we investigated the bifurcations occurring across the transition from monophasic to polyphasic sleep behavior motivated by  changes that occur in the time constants of the homeostatic sleep drive during development.  In our model, we introduced a scaling parameter $\chi$ in order to consistently scale $\tau_{hw}$ and $\tau_{hs}$: 

\begin{equation}
h'=\frac{(H_{max}-h)}{\chi \  \tau_{hw}}H[F_W-\theta_W]-\frac{h} {\chi \ \tau_{hs}}H[\theta_W-F_W]
\end{equation}

\noindent
with $0 < \chi$. Note that in our notation, $\chi$ represents a scaling factor on the time constants  $\tau_{hw}$ and $\tau_{hs}$ although other authors have used $\chi$ to denote the time constants themselves \cite{PR, Skeldon}.  By scaling the time constants governing growth and decay of $h(t)$, we alter the rates at which the homeostatic sleep drive accumulates and dissipates in the system.  In addition to direct effects on $h$, this alters the interactions between $h$ and the circadian drive.

\begin{figure}[h!]
\centering
\includegraphics[width=\textwidth]{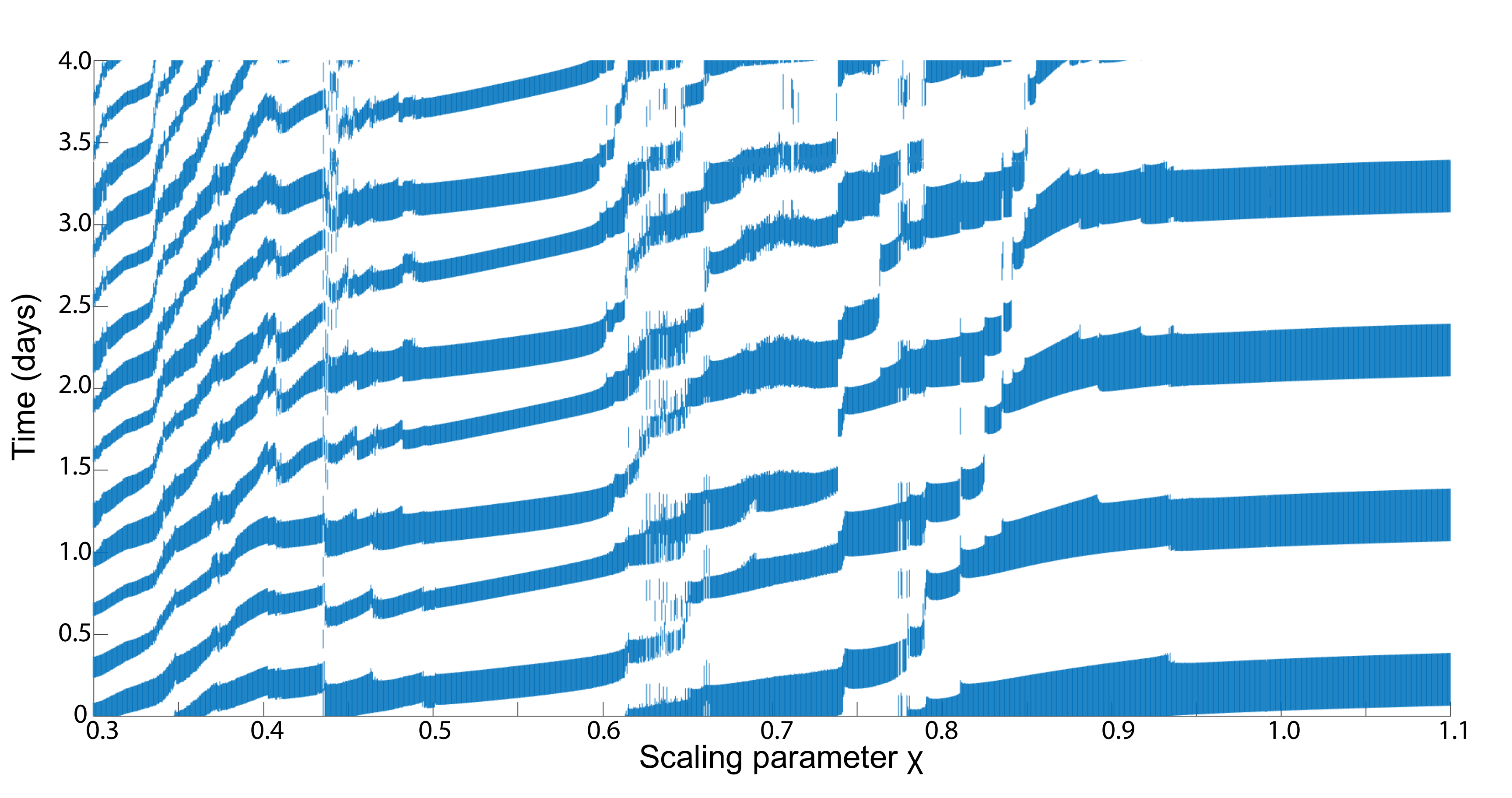}
\caption{Patterning of sleep-wake behavior varies with scaling parameter $\chi$.  Sleep periods over 4 days are shown as a function of $\chi$.  As $\chi$ decreases, the pattern of sleep transitions from one sleep cycle per day near $\chi=1$ to two sleep cycles per day near $\chi=0.55$ to three sleep cycles per day near $\chi=0.32$. The transitions increased in increments of 1 as $\chi$ decreased.}
\label{fig2}
\end{figure}

Using numerical simulations, we found that the number of sleep cycles per day appeared to increase in increments of one as $\chi$ decreased from 1.  To summarize these changes, we computed the time spent in sleep over a four day period as $\chi$ ranged between 0.3 and 1.1 (Figure \ref{fig2}). For certain intervals of $\chi$ values, the model produced a fixed number of sleep cycles per day.  For example, for values of $\chi$ with $0.866 < \chi < 1.9885$, one sleep cycle per day occurred; for $0.454 < \chi < 0.5865$, two sleep cycles per day occurred; and for $0.3085 < \chi < 0.331$ three sleep cycles per day occurred.  However, between the intervals of $\chi$ values corresponding to a single number of sleep cycles per day, there were $\chi$ values which produced variable numbers of sleep cycles per day.  For example, for $\chi = 0.72$ the model alternated between one and two sleep cycles per day. 


\begin{figure}
\centering
\includegraphics[width=\textwidth]{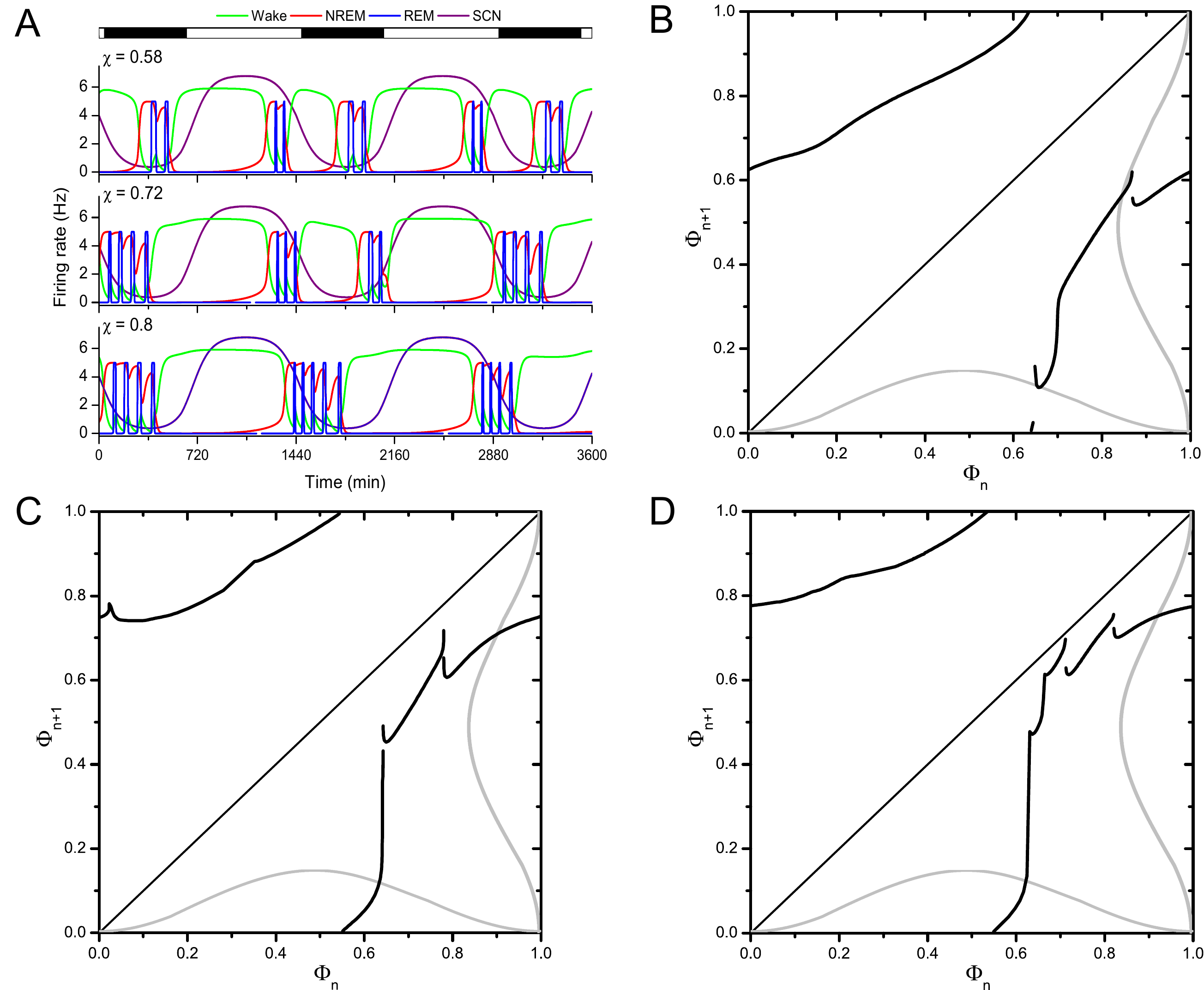}
\caption{Time traces and maps reflect changes in sleep-wake architecture as $\chi$ varies.  A.  Time traces show the number of sleep bouts per day changing with $\chi$.  For $\chi=0.8$ (bottom), one sleep cycle occurs each day although sleep cycles contain variable numbers of REM bouts.  For $\chi=0.72$ (middle), simulated behavior alternates between one and two sleep cycles per day.  For $\chi=0.58$ (top), two sleep cycles occur every day.  B.  Map for $\chi=0.58$ does not have any fixed points.  C.  Map for $\chi=0.72$ shows branches moving away from $y=x$ line. D.  Map for $\chi=0.8$ shows the configuration of branches just after the fixed point is lost.  }
\label{fig3}
\end{figure}

To better understand the underlying dynamics associated with the change in the number of sleep cycles per day, we examined the one-dimensional maps for the sleep-wake model for different values of $\chi$ (Figure \ref{fig3}).   Using the approach described in Section \ref{1Dmap}, we computed maps for values of $\chi < 1$ in order to track changes in the maps and the existence and stability of fixed points that corresponded to changes in the observed sleep-wake dynamics.  As $\chi$ decreases, the branch of the map associated with the stable steady state moves down until, for $\chi \approx 0.8$ the map no longer intersects the line $y=x$ (Figure \ref{fig3}D).  Thus, the fixed point of the map is lost as a direct result of the discontinuity, suggesting the occurrence of a border collision bifurcation.  Interestingly, the slope of the map curve at the fixed point undergoes changes in its sign and magnitude before the fixed point is lost at the cusp at the end of the map branch.  Specifically, the slope at the cusp is greater than one, so the fixed point loses stability before it loses existence.  This change in stability makes it difficult to relate the bifurcation in our system to established normal forms of piecewise smooth maps \cite{diBernardo2008, Granados}.  However, as described below, additional numerical evidence demonstrating period adding-like behavior in the system as $\chi$ changes suggests that the loss of the fixed point shares properties of a border collision bifurcation.

\begin{figure}[h!]
\centering
\includegraphics[width=\textwidth]{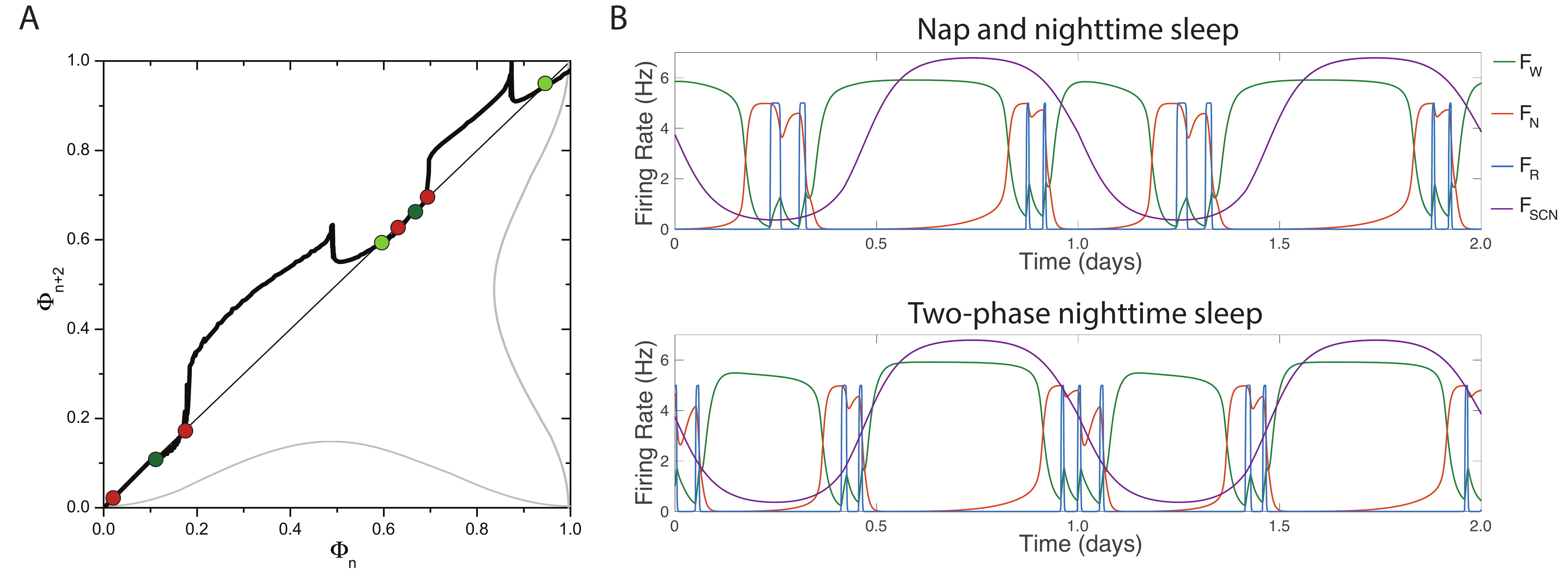}
\caption{For $\chi=0.58$, stable 2-cycles are associated with patterns of two sleep periods per day.  A.  The 2nd-return map shows 4 stable (light and dark green) and 4 unstable (red) period 2 points.  B. The light green points correspond to a pattern of nap and nighttime sleep.  C.  The dark green points correspond to a pattern of two-phase nighttime sleep.}
\label{fig4}
\end{figure}

Following the loss of the fixed point near $\chi=0.8$, the full model produces behavior in which some days exhibit one sleep cycle and other days exhibit two sleep cycles.  As $\chi$ decreases further, the map continues to evolve in ways that do not permit the re-emergence of a fixed point corresponding to a single, entrained sleep period. However, when $\chi=0.58$ (Figure \ref{fig3}B), the model produces a stable pattern of two sleep cycles per day. This behavior corresponds to the existence of a stable two-cycle for the map corresponding to fixed points of the second return map satisfying $\Phi_{n+2}=\Phi_n$ (Figure \ref{fig4}A).  Consideration of second return maps associated with $\chi$ values in this range show that there are actually two stable two-cycles present; bistability occurs for $0.5645<\chi<0.5865.$  As seen in the map for $\chi=0.58$ the stable two-cycles correspond to distinct behaviors:  one represents a daytime nap and a period of nighttime sleep and one represents bi-phasic sleep occuring at the beginning and end of the dark period (Figure \ref{fig4}B).  The large vertical discontinuity that is present in the first return map is absent in the second return map.  For increasing $\chi$, the loss of the stable two-cycles occurs for $\chi \approx 0.5865$ in saddle-node bifurcations.  For decreasing $\chi$, the two-cycle associated with the two-phase dark period sleep behavior loses stability for $\chi \approx 0.5645$, and the two-cycle associated with the nap and nighttime sleep loses stability for $\chi \approx 0.48$.  Thus, although the two-cycle associated with nap and nighttime sleep is stable over a much larger range of $\chi$ values compared to the two-cycle associated with two-phase nighttime sleep, both behaviors may coexist in the model.

It is important to note that a stable pattern of $n$ sleep cycles per day may not correspond to a stable $n$-cycle of the map.  For example, for $\chi=0.9$, one sleep cycle occurs per day, but the fixed point of the map is not stable.  Although the dynamics of the system produce one sleep cycle per day, the time of sleep onset varies slightly between days, resulting in sleep onset at different phases on different days.  This variability is reflected in the regions of Figure \ref{fig2} in which the pattern of sleep and wake does not repeat periodically across the 4 day period, and it is related to the bifurcation in the number of REM bouts per sleep cycle which will be discussed further in Section \ref{REMBifns}.

\subsection{Bifurcation structure in number of sleep cycles per day}
\begin{figure}[h!]
\centering
\includegraphics[width=\textwidth]{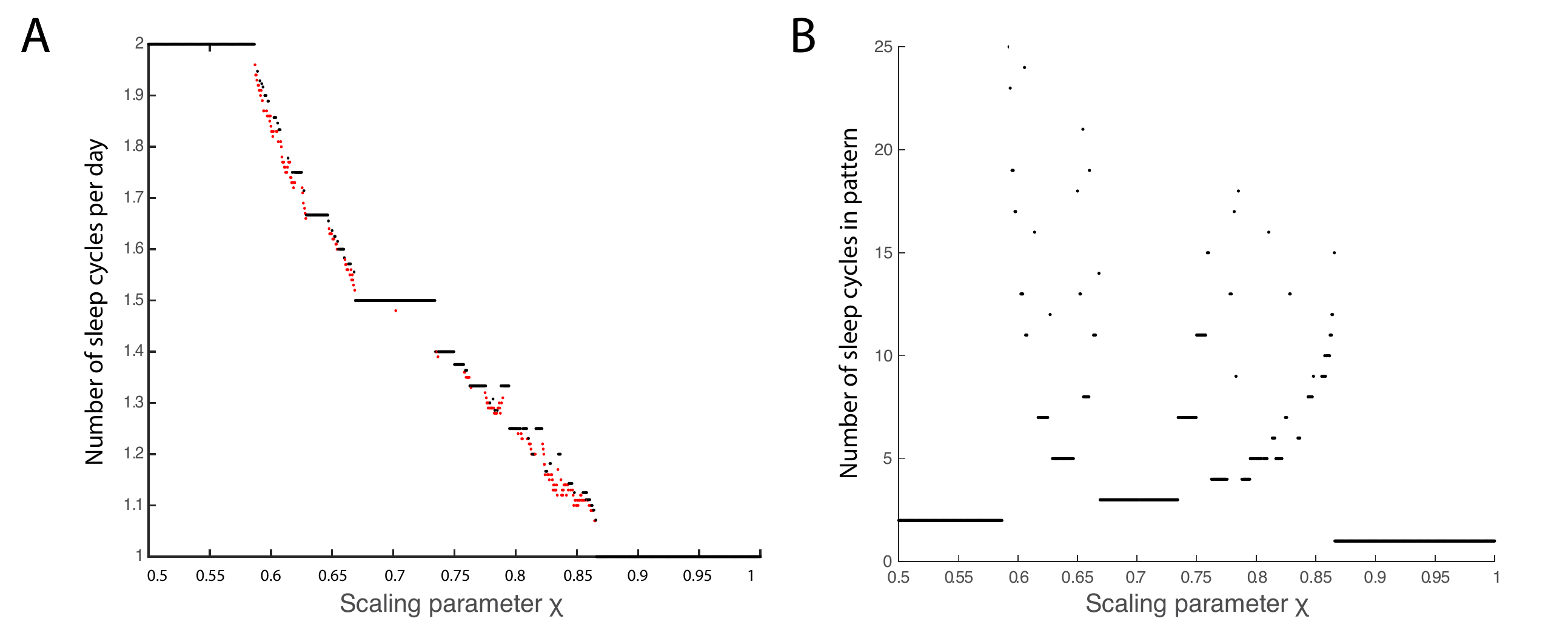}
\caption{Transition from monophasic to biphasic sleep patterns as $\chi$ is varied quantified by the number of sleep cycles per day (A) and the number of sleep cycles in the stable sleep pattern (B).  A:  For stable patterns, the sum of the number of sleep cycles in the stable repeating pattern divided by the number of days in the pattern is plotted (black dots); for quasi-periodic patterns, the total number of sleep cycles divided by the total number of days in a simulation lasting 2400 h is plotted (red dots). B: The number of sleep cycles in the stable patterns form an approximate bowl-shape as a function of $\chi$ typical of period adding bifurcations.  }
\label{fig5}
\end{figure}

To numerically investigate the structure of the bifurcations in the number of sleep cycles per day, we used two different methods to track the number of sleep cycles produced per day as $\chi$ varied across the transition region from one stable sleep cycle per day to two stable sleep cycles per day.  For a fixed value of $\chi$, the pattern of sleep cycles per day is summarized as a sequence $\{s_i \}$ where $s_i$ is the number of sleep cycles on day $i$.  For example, when $\chi=1$, one sleep cycle occurs every day and corresponds to the sequence $\{1,1,1,\ldots\} = \{1\}^{\infty}$.  Similarly, for $\chi=0.5$, two sleep cycles occur every day and are associated with the sequence $\{2,2,2,\ldots\} = \{2\}^{\infty}$.  In one method, we developed a pattern detection algorithm to numerically detect the sequence representing the number of sleep cycles per day for a given value of $\chi$ (see Supplementary Material for a description of the algorithm).  In a second method, we detected periodicity in the phases of sleep onsets and computed the number of sleep onsets and days in the periodic cycle.  We applied both methods to study the structure in the changes in the number of sleep cycles per day across the transition region from $\chi=1$ to $\chi=0.5$ and obtained qualitatively similar results.  The results shown in Figure \ref{fig5} are from the pattern detection algorithm.

For each value of $\chi$ from 0.5 to 1 with step size $5\times 10^{-4}$, we calculated the number of sleep cycles per day, $\rho$, which is similar to a rotation number (Figure \ref{fig5}A). 
For $\chi$ values at which a stable pattern was detected (black dots), $\rho$ was calculated as the  number of sleep cycles in the stable repeating pattern divided by the number of days in the pattern.  For example, the stable patterns $\{1\}^{\infty}$, $\{1,2\}^{\infty}$, and $\{2\}^{\infty}$ correspond to  $\rho = 1,$ 1.5, and 2, respectively.   
At values of $\chi$ for which no stable pattern was detected (red dots), an average $\rho$ was computed as the total number of sleep cycles divided by the total number of days in a simulation lasting 2400 h.  
    

 Overall, we found that stable patterns display some properties of a period adding bifurcation structure in rotation number and in number of sleep cycles (Figure \ref{fig5}B) with some major exceptions.  The rotation numbers of stable patterns between $1.5 \leq \rho \leq 2$, corresponding to $0.7 \geq \chi \geq 0.5$, occur at values $\rho=(2n+1)/(n+1)$ corresponding to patterns of the form $\{1, 2^n \}$ representing $n$ days of two sleep cycles followed by one day of one sleep cycle.  The rotation numbers increase monotonically and, for successive values of $n$, the associated values of $\rho$, $\frac{a}{b}$ and $\frac{c}{d}$,  form a pair of Farey neighbors that satisfy $|ad-bc|=1$, typical of period-adding bifurcations \cite{Granados}. 
 These solutions form the left-most border of the bowl-shape, typically observed in period adding bifurcations \cite{Granados}, in the plot of the number of sleep cycles in the stable patterns as a function of $\chi$ (Figure \ref{fig5}B).   However, rotation numbers for stable patterns did not exist continuously as $\chi$ was varied since we detected quasi-periodic patterns (Figure \ref{fig5}A, red dots) at $\chi$ values between the majority of intervals where stable patterns existed.  In a strict period adding bifurcation, the plot of stable rotation number as a function of $\chi$ forms a devil's staircase, defined as a continuous and monotonic  function which is constant locally almost everywhere \cite{Granados}.  By contrast, our plot of stable rotation number as a function of $\chi$ does not form a perfect devil's staircase, although an underlying devil's staircase-like structure is evident.  Average rotation numbers for the quasi-periodic solutions were approximately similar to neighboring stable rotation numbers reflecting patterns with similar numbers of sleep cycles per day but without strict periodicity.

%
For $0.7 \leq \chi \leq 1$, rotation numbers of stable patterns between $1.5 \geq \rho \geq 1$ occur at values $\rho=(n+2)/(n+1)$ for positive integers $n$ corresponding to patterns of the form $\{1^n, 2 \}^{\infty}$ representing $n$ days with one sleep cycle followed by one day with two sleep cycles.  In this range of $\chi$ values, stable rotation numbers did not decrease monotonically.  For example, stable solutions with rotation number $\rho = 4/3$ corresponding to the sleep pattern $\{1^2, 2\}^{\infty}$ occur over 2 intervals of $\chi$ values, $\approx (0.764,0.775)$ and $\approx (0.787, 0.794)$. 

Interestingly, the sleep patterns corresponding to these disjoint intervals differ in the numbers of REM sleep bouts that occur during the sleep episodes. To denote both sleep cycles per day and REM bouts per sleep cycle in patterns associated with particular values of $\chi$, we introduce the notation $\{n, n, (m, m) \}$ to denote $n$ REM bouts per sleep cycle on days with one sleep cycle per day and $m$ REM bouts per sleep cycle on days with two sleep cycles per day.  In the lower $\chi$ interval, the REM sleep pattern is $\{4, 4, (3, 3)\}^{\infty}$ representing 4 REM bouts during each sleep episode with 1 sleep cycle per day, and 3 REM bouts per sleep episode when there are 2 sleep cycles per day.  In the higher $\chi$ interval, the REM sleep pattern is $\{4, 3, (3, 4)\}^{\infty}$.  In the gap between the stable $\{1^2, 2\}^{\infty}$ patterns, stable patterns of the form $\{ (1, 1, 2)^m, (1, 1, 1, 2)^n \}^{\infty}$ ($n,m$ positive integers) exist as well as quasi-periodic solutions.  Stable sleep patterns with rotation numbers $\rho = 1.25$ corresponding to a $\{1^3, 2\}^{\infty}$ sleep pattern, and $\rho = 1.2$ corresponding to a $\{1^4, 2\}^{\infty}$ pattern also exist over disjoint $\chi$ intervals with different REM sleep patterning in the two intervals.  This non-monotonicity in the variation of rotation numbers for stable sleep patterns also deviates from a strict period adding bifurcation.  
     
In addition to the patterns of the forms $\{1^n, 2 \}^{\infty}$ and $\{1, 2^n \}^{\infty}$, other stable patterns of sleep cycles were observed for values of $\chi$ in the interval $[0.5, 1]$.  For example, the stable solutions with $\rho=8/5=1.6$ sleep cycles per day correspond to a pattern of 
\begin{equation}
\{(1,2),(1,2,2)\}^{\infty};
\end{equation}
stable solutions with $\rho=11/7\approx1.5714$ sleep cycles per day correspond to a pattern of
\begin{equation}
\{(1,2),(1,2),(1,2,2)\}^{\infty};
\end{equation}
and stable solutions with  $\rho=7/5=1.4$ sleep cycles per day correspond to a pattern of
\begin{equation}
\{(1,2),(1,1,2)\}^{\infty}.
\end{equation}
These more complex patterns are typical of period adding bifurcations as they exist for $\chi$ values between those exhibiting stable $\{1,2^n\}^{\infty}$ and $\{1,2^{n+1}\}^{\infty}$ patterns, or between $\{1^{n+1},2\}^{\infty}$  and $\{1^n,2\}^{\infty}$ patterns.  They are concatenations of those patterns and their rotation numbers are Farey neighbors.   

\section{Bifurcations in the number of REM bouts as dynamics of the homeostatic sleep drive vary}
\label{REMBifns}

As observed in the above results, the NREM-REM structure within sleep episodes also changes with $\chi$.
The resulting interaction between the number of sleep cycles per day and the REM structure of the sleep cycles contributed to some of the complexity observed in the $\chi$-dependence of sleep-wake patterning in our model compared to that reported in models of sleep-wake regulation that only describe two states \cite{Nakao, Phillips2013, Skeldon}.  To investigate the role of NREM-REM cycling on transitions in the number of sleep cycles per day, we quantified changes in REM sleep patterning and related $\chi-$dependent changes in REM sleep patterns to the associated maps.   

\subsection{NREM-REM cycling in single daily sleep episodes}

%
For this analysis, we focus on the bifurcation in the number of REM bouts per sleep cycle that takes place in the parameter regime preceding the bifurcation away from a single stable sleep cycle per day.  With the original homeostatic time constants, there is a single, consolidated sleep period during which four REM bouts occur as a result of ultradian cycling during sleep.  As described in Section \ref{3StateBifns}, the system continues to produce a single, consolidated sleep period as $\chi$ decreases from 1 to a value below 0.9, however, the fixed point that is associated with this single sleep period loses stability before it loses existence.  For values of $\chi$ around 0.9, the model produces a single sleep cycle per day, but the number of REM bouts varies in each sleep cycle.  This affects the duration of the sleep period and results in day-to-day variability in sleep onset, leading to the loss of stability of the fixed point in the associated map.  This variability in the phase of sleep onset is reflected in Figure \ref{fig2} in the lack of smoothness and consistency in the blue bands corresponding to sleep periods on different days for $\chi$ near 0.9.

\begin{figure}
\centering
\includegraphics[width=\textwidth]{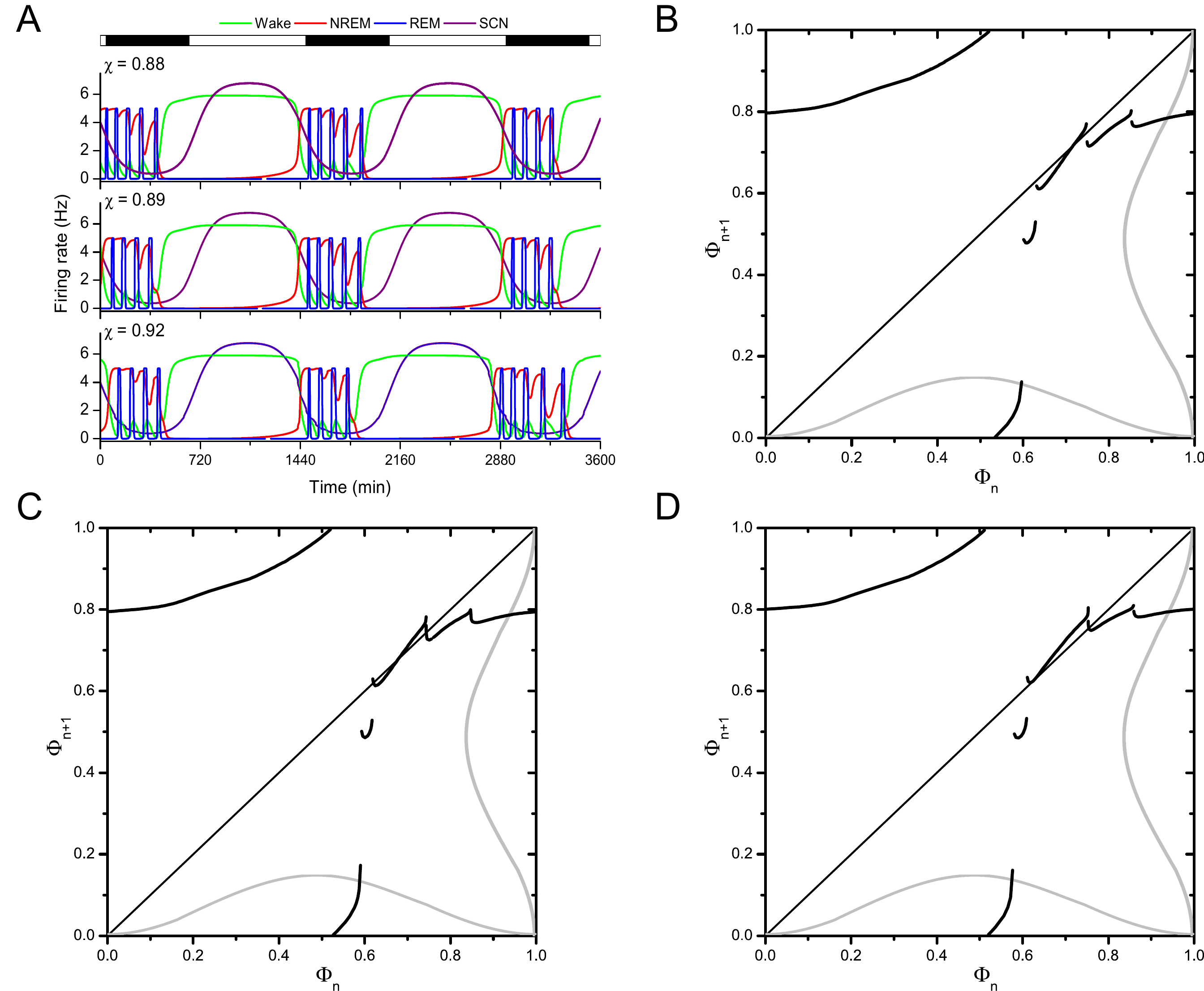}
\caption{Time traces and maps reflect changes in REM sleep patterning as $\chi$ varies.  A: Time traces show different patterns of REM bouts during a single nighttime sleep bout as $\chi$ changes.  For $\chi=0.88$ (top), REM bouts occur in a $\{5, 5, 4\}^{\infty}$ pattern.  For $\chi=0.89$ (middle), the number of REM bouts alternates in a $\{5, 4, 5, 4\}^{\infty}$ pattern.  For $\chi=0.92$ (bottom), REM bouts occur in a $\{4, 4, 5\}^{\infty}$ pattern. B-D: In the associated maps, when $\chi=0.88$ (B), $\chi=0.89$ (C), or $\chi=0.92$ (D), the fixed point is unstable since different numbers of REM bouts produce different phases of sleep onset on subsequent days. }
\label{fig6}
\end{figure}

In the map associated with the original homeostatic time constants, each of the branches of the piecewise continuous map is associated with a fixed number of REM bouts per sleep cycle \cite{BXandDB}.  The stable fixed point lies on a branch corresponding to four REM bouts per sleep cycle, and the branch to the left of the stable fixed point corresponds to five REM bouts per sleep cycle.  As $\chi$ decreases, the fixed point moves on to the cusp that lies between these two branches of the map (Figure \ref{fig6}B-D).  Since the magnitude of the slope is greater than one on this cusp, the fixed point loses stability.  However, the trajectory moves between these branches in a higher order cycle, thereby producing different numbers of REM bouts in successive sleep periods:  some sleep cycles have four REM bouts while others have five.  This variability in the structure within the sleep cycle leads to variability in the phase of sleep onset, consistent with the loss of stability of the fixed point for values of $\chi$ near 0.9.

For example, as seen in the model time traces corresponding to $\chi$ near 0.9 (Figure \ref{fig6}A), a parameter value of $\chi=0.92$ (bottom panel) produces a single sleep cycle per day, however, the number of REM bouts per sleep period varies in a $\{4, 4, 5\}^{\infty}$ pattern. As $\chi$ decreases to $0.89$ (middle panel) and $0.88$ (top panel), the system alternates between four and five REM bouts in the nighttime sleep period in a $\{4, 5\}^{\infty}$ pattern and in a $\{5, 5, 4\}^{\infty}$ pattern of REM bouts, respectively.  The appearance of sleep episodes with five REM bouts is associated with the appearance of unstable fixed points on the map branch corresponding to five REM bouts per sleep cycle for these values of $\chi$ (Figure \ref{fig6}B-D).   Due to the cusp structure of the map, for decreasing $\chi$, these fixed points associated with five REM bouts per sleep episode lose existence before they gain stability.  Thus, there is no value of $\chi$ that produces a stable daily pattern consisting of a single sleep cycle with five REM bouts.  

\subsection{Bifurcation structure in the number of REM bouts per sleep period}

\begin{figure}
\centering
\includegraphics[width=\textwidth]{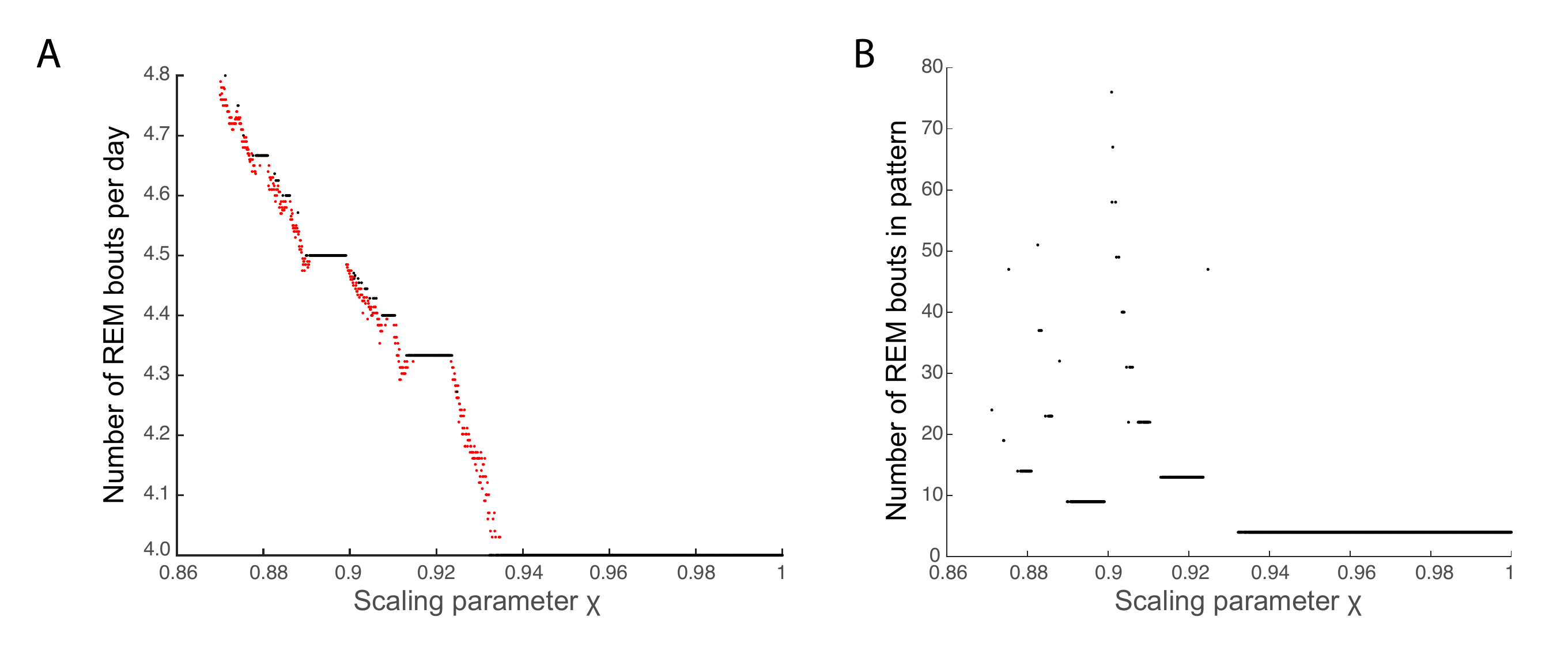}
\caption{Bifurcation structure as $\chi$ is varied in the number of REM bouts per sleep episode in solutions with one sleep cycle per day. A:  For stable solutions, the number of REM bouts per sleep cycle was computed by dividing the total number of REM bouts in the stable pattern by the number of sleep cycles in that pattern (black dots).  For unstable solutions, the average number of REM bouts per sleep cycle was approximated by dividing the number of REM bouts in the entire simulation by the number of sleep cycles in the simulation (red dots).  B: The number of REM bouts in the stable patterns form a truncated bowl-shape as $\chi$ is varied reflecting the truncated period adding bifurcation structure.}
\label{fig7}
\end{figure}

To investigate the change in the pattern of REM bouts per sleep cycle as $\chi$ decreases, we used an approach similar to that described for detecting the structure in the change of sleep cycles per day.  Specifically, we applied the pattern detection algorithm (see Supplementary Material) to the number of REM bouts per sleep cycle and computed, $\rho_{REM}$, the average number of REM bouts per sleep cycle by dividing the total number of REM bouts in a stable pattern by the number of sleep cycles in that pattern.  If no pattern was detected, the average number of REM bouts per sleep cycle was approximated by dividing the number of REM bouts in the entire simulation by the number of sleep cycles in the simulation.  In Figure \ref{fig7}, $\rho_{REM}$ values for stable solutions are plotted as black points and those for quasi-periodic solutions are red points.

For $\chi=1$, $\rho_{REM}=4$, and $\rho_{REM}$ increases as $\chi$ decreases reflecting the appearance of sleep episodes with five REM bouts.  In a period adding bifurcation, we would expect to see stable solutions with REM patterning of the form $\{4^n, 5\}$ and $\{4, 5^n\}$ for positive integers $n$.  Our model displays only a limited number of these stable solutions, namely $\{4, 5\}^{\infty}$ with $\rho_{REM}=4.5$, $\{4^2, 5\}^{\infty}$ with $\rho_{REM}=4.333$ and $\{4, 5^k\}^{\infty}$ for $k = 2 - 4$ corresponding to $\rho_{REM}=4.666, 4.75$ and $4.8$.  The other stable solutions are concatenations of two of these patterns that exist in the $\chi$ interval between the patterns.  For example, between the stable solutions with $\rho_{REM}=4.333$ and $4.5$ are stable patterns of the form $\{(4, 4, 5), (4, 5)^n\}^{\infty}$ where $n$ increases as $\chi$ decreases.  These solutions are evident in the vertical cascade of stable solutions in Figure \ref{fig7}B between the stable solutions with 13 and 9 REM bouts that increase in increments of 9 REM bouts as $\chi$ decreases.  So while in this $\chi$ interval period adding bifurcation structure is intact, model solutions display only a portion of the full period adding behavior.  In applications, it is common to encounter situations like this in which a period adding bifurcation scenario is truncated \cite{Avrutin2013}.  

In $\chi$ intervals between the stable solutions, the model displayed quasi-periodic behavior in the number of REM bouts per sleep cycle (red dots in Figure \ref{fig7}B).  In these solutions, each sleep episode had either four or five REM bouts, but the sequence of number of REM bouts per sleep cycle did not repeat.  The existence of such quasi-periodic solutions deviates from a strict period adding bifurcation structure in the number of REM bouts per sleep cycle.

\section{Discussion}
\label{discussion}

In this study, we presented a computationally-based analysis of the bifurcations produced by varying homeostatic time constants in a sleep-wake network model that simulates three states: wake, NREM sleep, and REM sleep.  These bifurcations were investigated using an associated piecewise continuous one-dimensional map.  The map was constructed from the high-dimensional, physiologically-based sleep-wake regulatory network model for human sleep, and transitions in the numbers of average simulated sleep cycles per day could be understood as a border collision bifurcation.  Specifically, decreasing the homeostatic time constants in the model caused the stable fixed point corresponding to entrained monophasic sleep-wake behavior to move through the discontinuity of the map.  Relatedly, transitions in the number of REM bouts per sleep cycle could be understood in terms of a loss of stability of this fixed point prior to the loss of existence.  The loss of stability was directly related to the non-monotonicity of the map at the cusp points separating different branches of the map.  

Considering the average number of sleep cycles as a function of the scaling parameter, $\chi$, revealed a structure with some properties of a period-adding bifurcation, but also some significant differences.  In the transition from monophasic to stable biphasic sleep patterns, we found $\chi$ intervals in which the numbers of sleep cycles per day exhibited the stable repeating patterns expected in a period adding bifurcation. However, these $\chi$ intervals were not contiguous and in some cases were not continuous.  Additionally, quasi-periodic solutions existed in $\chi$ intervals between the stable solutions.  Interestingly, focusing on the dynamics of the model for values of $\chi$ before the bifurcation point associated with the end of stable, monophasic sleep revealed a truncated period adding-like structure in the average number of REM bouts per sleep cycle. Again, quasi-periodic solutions existed in $\chi$ intervals between stable solutions. This suggests a type of nested hierarchy of bifurcations taking place in sleep patterns as the time constants of the homeostatic sleep drive decrease.  We also identified a range of $\chi$ values associated with bistability between two cycles corresponding to distinct sleep behaviors:  a nap-nighttime sleep pattern and a two-phase nighttime sleep pattern.  Although the nap-nighttime sleep pattern is more familiar in modern life and corresponds to sleep in early childhood as well as siesta cultures, two-phase nighttime sleep has been observed under short photoperiods and may represent historical sleep patterns that preceded modern artificial lighting conditions \cite{Wehr}. 

Our results also point to a role for REM sleep in the stabilization and de-stabilization of sleep patterns across the transition from monophasic to biphasic sleep.  Sleep patterns of the form $\{1^k, 2\}^{\infty}$ for $k = 2, 3$ and $4$ were stable in two disjoint $\chi$ intervals and the difference between the solutions in the two $\chi$ intervals were in the number of REM bouts per sleep episode.  In our model, a change in the number of REM bouts leads to a finite change in the duration of the sleep episode due to the fact that REM bout durations are governed by separate dynamical mechanisms than sleep episode durations.  Additionally, the propensity for REM bouts varies with the circadian rhythm \cite{GDBandB}. Thus, as $\chi$ varies, introducing a graded change in sleep episode durations as well as shifting the phases of the circadian rhythm over which sleep occurs, the number of REM bouts that can ``fit" in each sleep episode varies, leading to different stable sequences in the number of REM bouts per sleep episode but the same overall pattern of sleep cycles per day.

We previously determined that the interaction of the time scales associated with the homeostatic sleep drive and the circadian rhythm plays a key role in determining the behavior produced by the network \cite{ BXandDB, DBandB_SIADS2012}.  We exploited the slower time scales of the sleep homeostat and the circadian drive to apply fast-slow decomposition principles to understand underlying model dynamics.  For an appropriate range of fixed values for circadian variables, we identified a Z-shaped folded surface around which trajectories moved as they transitioned from wake to sleep over a typical 24 h cycle.  In this analysis, we noted that the separation of time scales that is necessary for a rigorous fast-slow decomposition was not always preserved in this system.  Thus, interactions among the time scales of the homeostatic sleep drive and the circadian variables could result in behavior that was not predicted by the system associated with the singular limit in which circadian variables were taken to be fixed parameters.  This interaction of time scales also complicated the construction of the one-dimensional map.  In general, the map was computed from initial conditions taken as the knee of the Z-shaped bifurcation diagram associated with a fixed value of the circadian variables.  However, for some regions, these initial conditions failed to produce a rapid transition from wake to sleep \cite{BXandDB}.  At these points, the change in circadian variables, and hence the change in the location of the knee of the Z-shaped bifurcation diagram, outpaced the change in the homeostatic sleep drive, thereby preventing a transition to sleep.  Biologically, this interaction may explain the existence of a ``wake maintenance zone'' for sleep that occurs in the morning approximately 5 hours after minimum core body temperature \cite{Strogatz}.  However, in the map, this interaction produced horizontal and vertical discontinuities in the piecewise continuous one-dimensional map generated from initial conditions at the saddle-node point.  Computation of the map for circadian phases associated with the horizontal discontinuity required a choice of initial conditions on the unstable manifold of the saddle-node point \cite{BXandDB}.  In this work, we focused on applying the one-dimensional map to understand bifurcations of the full model.  As the scaling parameter $\chi$ decreased, the discontinuities in the map changed but generally persisted.  This likely reflects changes in the interaction between the dynamics of the circadian and homeostatic sleep drives over 24 h since $\chi$ directly alters the time constants of the homeostatic sleep drive.  Interestingly, for $\chi=0.58$, the second return map does not show any discontinuities suggesting that, with these time constants, the homeostatic sleep drive is never outpaced by the circadian variables.

A key motivation for this work was to improve understanding of the effects of increasing the rates of growth and decay of homeostatic sleep need in a three-state model of sleep-wake behavior.  We found that such a change resulted in a transition from monophasic to polyphasic sleep.  This is consistent with  characterization of the time constants of the homeostat in multiple species \cite{Phillips2013,Tobler} and in different human life stages \cite{Acebo2005, Iglowstein, Jenni2006} that suggests more frequent sleep periods are associated with increased build up of sleep need.  However, REM sleep timing, duration, and architecture also vary in different species and in humans at different life stages \cite{Ohayon, SiegelChap8, Tobler1995}.  The role of REM sleep in complicating the dynamics associated with the model's transition from monophasic to biphasic sleep suggest that REM-NREM sleep architecture should be considered when evaluating dynamics of monophasic and polyphasic sleep.


Similar bifurcations structures have been observed in other models with slow variables \cite{Phillips2013, Skeldon,Zhang}.   In mathematical models that describe sleep-wake behavior as two states, wake and sleep, without separately considering REM and NREM sleep, this dependence has been established for homeostatic time constants \cite{Phillips2013, Skeldon}. In particular, sleep patterns transition from monophasic to polyphasic as the time constants for the homeostatic sleep drive descrease in the Two Process model, a classic phenomenological model for the circadian regulation of sleep and wake states, and in a physiologically-based sleep-wake flip-flop model \cite{PR}.  Under baseline conditions associated with adult humans, similarities in underlying dynamics of these sleep-wake models and the model considered here have been established \cite{BandDB2014, Skeldon}.  Indeed, it has been shown that in a specific parameter regime, the Two Process model is dynamically equivalent to the sleep-wake flip-flop model \cite{Skeldon}.  Further analysis of the Two Process model has shown that its dynamics are captured exactly by a one-dimensional circle map that is piecewise smooth \cite{Nakao, Skeldon}.  A bifurcation analysis of the Two Process model in this parameter regime revealed a strict period adding bifurcation structure in the number of sleep episodes per day as the time constant of the homeostatic sleep drive was varied and sleep patterns varied from monophasic to polyphasic \cite{Skeldon}.  Our results indicate that REM sleep dynamics add complexity to the bifurcation structure suggesting that there may be important aspects of the transition from polyphasic to monophasic sleep that are not captured by two-state models.  Future work characterizing homeostatic time constants and sleep-wake behavior in preschoolers as they transition from napping to non-napping behavior is needed to determine the relative importance of accounting for REM sleep during this transition.



\section*{Acknowledgments}
We would like to thank Dr. Monique LeBourgeois for helpful conversations about sleep in early childhood.

\bibliographystyle{siamplain}
\bibliography{ThreeStateBifnPaper}

\begin{thebibliography}{10}

\bibitem{Abrahamson2001}
{\sc E.~E. Abrahamson, R.~K. Leak, and R.~Y. Moore}, {\em The suprachiasmatic
  nucleus projects to posterior hypothalamic arousal systems}, Neuroreport, 12
  (2001), pp.~435--440.

\bibitem{Acebo2005}
{\sc C.~Acebo, A.~Sadeh, R.~Seifer, O.~Tzischinsky, A.~Hafer, and
  M.~Carskadon}, {\em Sleep/wake patterns derived from activity monitoring and
  maternal report for healthy 1- to 5-year old children}, Sleep, 28 (2005),
  pp.~1568--1577.

\bibitem{Avrutin2013}
{\sc V.~Avrutin and I.~Sushko}, {\em A gallery of bifurcation scenarios in
  piecewise smooth 1d maps}, in Global Analysis of Dynamic Models in Economics
  and Finance, G.~e.~a. Bischi, ed., Springer-Verlag, 2013.

\bibitem{Basheer}
{\sc R.~Basheer, R.~Strecker, M.~Thakkar, and R.~McCarley}, {\em Adenosine and
  sleep-wake regulation}, Prog Neurobiol, 73 (2004), pp.~379--396.

\bibitem{BandDB2014}
{\sc V.~Booth and C.~G. Diniz~Behn}, {\em Physiologically-based modeling of
  sleep$/$wake regulatory networks}, Math Biosci, 250 (2014), pp.~54--68.

\bibitem{BXandDB}
{\sc V.~Booth, I.~Xique, and C.~G. Diniz~Behn}, {\em One-dimensional map for
  the circadian modulation of sleep in a sleep-wake regulatory network model
  for human sleep}, SIAM J App Dyn Systems,  (In Press).

\bibitem{CarskadonDementChap2}
{\sc M.~Carskadon and W.~Dement}, {\em Normal human sleep: an overview}, in
  Principles and Practice of Sleep Medicine, M.~Kryger, T.~Roth, and W.~Dement,
  eds., Elsevier Saunders, 2011.

\bibitem{CzeislerBuxtonChap35}
{\sc C.~A. Czeisler and O.~M. Buxton}, {\em The human circadian timing system
  and sleep$/$wake regulation}, in Principles and Practice of Sleep Medicine,
  M.~Kryger, T.~Roth, and W.~Dement, eds., Elsevier Saunders, 2011.

\bibitem{DayanAbbott}
{\sc P.~Dayan and L.~Abbott}, {\em Theoretical Neuroscience: Computational and
  Mathematical Modeling of Neural Systems}, The MIT Press, 2001.

\bibitem{Meijer}
{\sc T.~Deboer, V.~M., L.~Detarii, and J.~Meijer}, {\em Sleep states alter
  activity of suprachiasmatic nucleus neurons}, Nat Neurosci, 6 (2003),
  pp.~1086--1090.

\bibitem{Deco2008}
{\sc G.~Deco, V.~K. Jirsa, P.~A. Robinson, M.~Breakspear, and K.~Friston}, {\em
  The dynamic brain: from spiking neurons to neural masses and cortical
  fields}, PLoS Comput Biol, 4 (2008), p.~e1000092.

\bibitem{diBernardo2008}
{\sc M.~di~Bernardo, C.~Budd, A.~Champneys, and P.~Kowalczyk}, {\em
  Piecewise-smooth Dynamical Systems: Theory and Applications}, Springer, 2008.

\bibitem{DBAandB_SIADS2013}
{\sc C.~Diniz~Behn, A.~Ananthasubramaniam, and V.~Booth}, {\em Contrasting
  existence and robustness of {REM/NREM} cycling in physiologically based
  models of {REM} sleep regulatory networks}, SIAM J on App Dyn Systems, 12
  (2013), pp.~279--314.

\bibitem{DBandB_SIADS2012}
{\sc C.~Diniz~Behn and V.~Booth}, {\em A fast-slow analysis of the dynamics of
  {REM} sleep}, SIAM J on App Dyn Systems, 11 (2012), pp.~212--242.

\bibitem{Edgar1993}
{\sc D.~Edgar, W.~Dement, and C.~Fuller}, {\em Effect of {SCN} lesions on sleep
  in squirrel monkeys: evidence for opponent processes in sleep-wake
  regulation}, J Neurosci, 13 (1993), pp.~1065--1079.

\bibitem{Ermentrout1998}
{\sc G.~B. Ermentrout}, {\em Neural networks as spatio-temporal pattern-forming
  systems}, Rep. Prog. Phys., 61 (1998), pp.~353--430.

\bibitem{xpp}
{\sc G.~B. Ermentrout}, {\em Simulating, analyzing, and animating dynamical
  systems: a guide to XPPAUT for researchers and students}, Society for
  Industrial and Applied Mathematics, Philadelphia, 2002.

\bibitem{Feinberg}
{\sc I.~Feinberg and T.~Floyd}, {\em Systematic trends across the night in
  human sleep cycles}, Psychophysiology, 16 (1979), pp.~283--291.

\bibitem{Fleshner}
{\sc M.~Fleshner, V.~Booth, D.~Forger, and C.~Diniz~Behn}, {\em Circadian
  regulation of sleep-wake behavior in nocturnal rats requires multiple signals
  from suprachiasmatic nucleus}, Phil. Trans. R. Soc. A, 369 (2011),
  pp.~3855--3883.

\bibitem{Forger1999}
{\sc D.~B. Forger, M.~E. Jewett, and R.~E. Kronauer}, {\em A simpler model of
  the human circadian pacemaker}, J Biol Rhythms, 14 (1999), pp.~532--537.

\bibitem{Gaudreau2001}
{\sc H.~Gaudreau, J.~Carrier, and J.~Montplaisir}, {\em Age-related
  modifications of {NREM} sleep {EEG}: from childhood to middle age}, J Sleep
  Res, 10 (2001), pp.~165--172.

\bibitem{GDBandB}
{\sc R.~D. Gleit, C.~Diniz~Behn, and V.~Booth}, {\em Modeling interindividual
  differences in spontaneous internal desynchrony patterns}, J Biol Rhythms, 28
  (2013), pp.~339--355.

\bibitem{Granados}
{\sc A.~Granados, L.~Alseda, and M.~Krupa}, {\em The period adding and
  incrementing bifurcations: from rotation theory to applications}, SIAM
  Review, 59 (2017), pp.~225--292.

\bibitem{Huang}
{\sc Z.~Huang, Y.~Urade, and O.~Hayaishi}, {\em Prostaglandins and adenosine in
  the regulation of sleep and wakefulness}, Curr Opin Pharmacol, 7 (2007),
  pp.~33--38.

\bibitem{Iglowstein}
{\sc I.~Iglowstein, O.~Jenni, L.~Molinari, and R.~Largo}, {\em Sleep duration
  from infancy to adolescence: reference values and generational trends},
  Pediatrics, 111 (2003), pp.~302--307.

\bibitem{Jenni2005}
{\sc O.~Jenni, P.~Achermann, and M.~Carskadon}, {\em Homeostatic sleep
  regulation in adolescents}, SLEEP, 28 (2005), pp.~1446--1454.

\bibitem{Jenni2006}
{\sc O.~Jenni and M.~LeBourgeois}, {\em Understanding sleep–wake behavior and
  sleep disorders in children: the value of a model}, Curr Opin Psychiatry, 19
  (2006), pp.~282--287.

\bibitem{KaplanGlass}
{\sc D.~Kaplan and L.~Glass}, {\em Understanding Nonlinear Dynamics}, Springer
  Science Business Media, 2012.

\bibitem{KronauerErrata}
{\sc R.~E. Kronauer, D.~B. Forger, and M.~E. Jewett}, {\em Errata: Quantifying
  human circadian pacemaker response to brief, extended, and repeated light
  stimuli over the photopic range}, J Biol Rhythms, 15 (2000), pp.~184--186.

\bibitem{Kumar}
{\sc R.~Kumar, A.~Bose, and B.~Mallick}, {\em A mathematical model towards
  understanding the mechanism of neuronal regulation of wake-{NREMS}-{REMS}
  states}, PLoS One, 7 (2012), p.~e2059.

\bibitem{McCarleyHobson}
{\sc R.~W. McCarley and J.~A. Hobson}, {\em Neuronal excitability modulation
  over the sleep cycle: a structural and mathematical model}, Science, 189
  (1975), pp.~58--60.

\bibitem{McCarleyMassaquoi}
{\sc R.~W. McCarley and S.~G. Massaquoi}, {\em A limit-cycle mathematical-model
  of the {REM}-sleep oscillator system}, Am J Physiol, 251 (1986),
  pp.~R1011--R1029.

\bibitem{Mistlberger2005}
{\sc R.~E. Mistlberger}, {\em Circadian regulation of sleep in mammals: role of
  the suprachiasmatic nucleus}, Brain Res Rev, 49 (2005), pp.~429--454.

\bibitem{Nakao}
{\sc M.~Nakao, H.~Sakai, and M.~Yamamoto}, {\em An interpretation of the
  internal desynchronizations based on dynamics of the two-process model}, Meth
  Inform Med, 36 (1997), pp.~282--285.

\bibitem{Ohayon}
{\sc M.~Ohayon, M.~Carskadon, C.~Guilleminault, and M.~Vitiello}, {\em
  Meta-analysis of quantiative sleep parameters from childhood to old age in
  healthy individuals: developing normative sleep values across the human
  lifespan}, {SLEEP}, 27 (2004), pp.~1255--1273.

\bibitem{Phillips2013}
{\sc A.~J.~K. Phillips, B.~D. Fulcher, P.~A. Robinson, and E.~B. Klerman}, {\em
  Mammalian rest/activity patterns explained by physiologically based
  modeling}, PLoS Comput Biol, 9 (2013),
  \url{https://doi.org/10.1371/journal.pcbi.1003213}.

\bibitem{PR}
{\sc A.~J.~K. Phillips and P.~A. Robinson}, {\em A quantitative model of
  sleep-wake dynamics based on the physiology of the brainstem ascending
  arousal system}, J Biol Rhythms, 22 (2007), pp.~167--179.

\bibitem{Rempe}
{\sc M.~Rempe, J.~Best, and D.~Terman}, {\em A mathematical model of the
  sleep\/wake cycle}, J Math Biol, 60 (2010), pp.~615--644.

\bibitem{Rusterholz}
{\sc T.~Rusterholz, R.~Durr, and P.~Achermann}, {\em Inter-individual
  differences in the dynamics of sleep homeostasis}, SLEEP, 33 (2010),
  pp.~491--498.

\bibitem{Saper2001}
{\sc C.~B. Saper, T.~C. Chou, and T.~E. Scammell}, {\em The sleep switch:
  hypothalamic control of sleep and wakefulness}, Trends Neurosci, 24 (2001),
  pp.~726--731.

\bibitem{Saper2005}
{\sc C.~B. Saper, T.~E. Scammell, and J.~Lu}, {\em Hypothalamic regulation of
  sleep and circadian rhythms}, Nature, 437 (2005), pp.~1257--1263.

\bibitem{SerkhForger}
{\sc K.~Serkh and D.~B. Forger}, {\em Optimal schedules of light exposure for
  rapidly correcting circadian misalignment}, PLoS Comput Biol, 10 (2014),
  p.~e1003523.

\bibitem{SiegelChap8}
{\sc J.~Siegel}, {\em {REM} sleep}, in Principles and Practice of Sleep
  Medicine, M.~Kryger, T.~Roth, and W.~Dement, eds., Elsevier Saunders, 2011.

\bibitem{Skeldon}
{\sc A.~Skeldon, D.-J. Dijk, and G.~Derks}, {\em Mathematical models for
  sleep-wake dynamics: Comparison of the two-process model and a mutual
  inhibition neuronal model}, PLoS One, 9 (2014), p.~e103877.

\bibitem{Strogatz}
{\sc S.~Strogatz, R.~E. Kronauer, and C.~A. Czeisler}, {\em Circadian pacemaker
  interferes with sleep onset and specific times each day: role in insomnia},
  Am J Physiol - Reg, Int, and Comp Physiol, 253 (1987), pp.~R172--R178.

\bibitem{Tamakawa}
{\sc Y.~Tamakawa, A.~Karashima, Y.~Koyama, N.~Katayama, and M.~Nakao}, {\em A
  quartet neural system model orchestrating sleep and wakefulness mechanisms},
  Journal of Neurophysiology, epub (2006), pp.~2055--2069.

\bibitem{Tobler1995}
{\sc I.~Tobler}, {\em Is sleep fundamentally different between mammalian
  species?}, Behavioral Brain Research, 69 (1995), pp.~35--41.

\bibitem{Tobler}
{\sc I.~Tobler, P.~Franken, L.~Trachsel, and A.~Borbely}, {\em Models of sleep
  regulation in mammals}, Journal of Sleep Research, 1 (1992), pp.~125--127.

\bibitem{Wehr}
{\sc T.~A. Wehr}, {\em In short photoperiods, human sleep is biphasic}, J Sleep
  Res, 1 (1992), pp.~103--107.

\bibitem{WilsonCowan}
{\sc H.~R. Wilson and J.~D. Cowan}, {\em Excitatory and inhibitory interactions
  in localized populations of model neurons}, Biophys J, 12 (1972), pp.~1--24.

\bibitem{Zhang}
{\sc Y.~Zhang, A.~Bose, and F.~Nadim}, {\em The influence of the a-current on
  the dynamics of an oscillator-follower inhibitory network}, SIAM J Applied
  Dynamical Systems, 8 (2009), pp.~1564--1590.

\end{thebibliography}
\end{document}


\maketitle

\section{Supplemental details on model construction}

As described in the main text, we assume that the mean firing rate of a presynaptic population, $F_Y(t)$, induces instantaneous expression of  neurotransmitter concentration, $C_i(t)$ \cite{DBandB_SIADS2012}. 
The index $Y \in \{W, N, R, SCN\}$ for wake-, NREM-, and REM-promoting populations and the SCN, and the index $i \in \{ E, G, N, A\}$ reflects the neurotransmitter associated with neuronal population $Y$: $i=E$ (noradrenaline) for $F_{YE}=F_W$; $i=G$ (GABA) for $F_{YG}=F_N$; and $i=A$ (acetylcholine) for $F_{YA}=F_R$.

The neurotransmitter concentration is set equal to its steady state response: $C_i(t)=C_{i\infty}(F_{Yi})$ where $C_{i\infty}(f)=\tanh(f/{\gamma_i})$ with the parameter $\gamma_i$ controlling the sensitivity of release as a function of presynaptic firing rate $f$.  


The mean post-synaptic firing rate $F_X(t)$ (in Hz) is governed by the equation
\begin{equation}
F_X'=\frac{F_{X\infty}(\sum_i g_{i,X}C_{i\infty}(F_{Yi}))-F_X}{\tau_X}
\label{FXsupp}
\end{equation}
where the steady state firing rate response function has the sigmoidal profile utilized in standard firing rate models \cite{WilsonCowan} (see reviews in \cite{Ermentrout1998,DayanAbbott,Deco2008}):
\begin{equation}
F_{X\infty}(z)=\frac{X_{max}}{2}(1+\tanh((z-\beta_X)/\alpha_X)).
\label{FXinftysupp}
\end{equation}
where the parameters $X_{max}$, $\alpha_X$, and $\beta_X$ represent the maximum firing rate, sensitivity of response, and half-activation threshold, respectively.  The argument  of $F_{X\infty}(\cdot)$ consists of the sum of effective synaptic inputs generated by the neurotransmitter concentrations released to population $X$.  The weighting parameters $g_{i,X}$ convert normalized concentrations $C_{i\infty}(\cdot)$ to effective synaptic input where the sign of each $g_{i,X}$ distinguishes between an excitatory ($g_{i,X} >0$) or an inhibitory ($g_{i,X}<0$) effect of the neurotransmitter.  The variable $F_X(t)$ evolves, with time constant $\tau_X$ in minutes, to the value determined by its steady state firing rate response function $F_{X\infty}(\cdot)$.  All parameter values for the sleep-wake model are listed in Table \ref{partable}.

\begin{table}
\begin{center}
\begin{tabular}{|| lr | lr | lr | lr ||}  \hline
$g_{A,W}$	& 0.8		& $g_{G,W}$	& -1.5		& $g_{S,W}$	& 0.5		& $g_{E,N}$	& -1.5		 \\ \hline
$g_{S,N}$		& -0.1		& $g_{A,R}$	& 2.2		& $g_{E,R}$	& -8		& $g_{G,R}$	& -1		\\  \hline
$g_{S,R}$		& 0.8		& & & & & &													\\ \hline
$\tau_W$		& 23		& $\tau_{N}$	& 10		& $\tau_{R}$	& 1		&$\tau_{SCN}$	& 0.5		\\ \hline
$\alpha_{W}$	& 0.4		& $\alpha_{N}$	& 0.2		& $\alpha_{R}$	& 0.1		&$\alpha_{SCN}$ & 0.7	\\ \hline
$\beta_{W}$	& -0.4	& $\beta_{SCN}$	& -0.1& 	& & & 								\\ \hline
$W_{max}$	& 6		& $N_{max}$	& 5		& $R_{max}$	& 5		& $SCN_{max}$	& 7	\\ \hline
$\gamma_{E}$ & 5		& $\gamma_G$	& 2.5		& $\gamma_A$	& 2.5		& $\gamma_S$		& 4	\\ \hline
$H_{max}$	& 323.88	&$\tau_{hw}$	& 946.8	& $\tau_{hs}$	& 202.2	& $\theta_W$		& 4 Hz	\\ \hline
$\tau_x$		& 24.2	&$\mu$		& 0.23	&$k$			& 0.55	&$\beta$		& 0.0075	 min$^{-1}$\\ \hline
$\alpha_0$	& 0.05 min$^{-1}$	&$I_0$		& 9500 lux	&$p$			& 0.5		& G 	& 33.75	 	\\ \hline
\end{tabular}
\end{center}
\caption{Parameter values for the reduced sleep-wake regulatory network model for human sleep.  For  $X = W$, $N$, $R$ and $SCN$ and $i=E$, $G$, $A$ and $S$, units for $\tau_X$ are minutes; units for $X_{max}$ and $\gamma_i$ are Hz; $\alpha_X$ and $\beta_X$ are in units of effective synaptic input and $g_{i,X}$ has units of (effective synaptic input / normalized concentration).  Units for $H_{max}$ are percentage mean SWA and units for  $\tau_{hw}$ and $\tau_{hs}$ are minutes.  For the circadian oscillator model, units for $\tau_x$ are hours and $p$, $\mu$, $k$, and $G$ are unitless. Remaining units are indicated in the table.}
\label{partable}
\end{table}

\section{Algorithm to detect pattern in number of sleep cycles per day}
    
  To examine the number of sleep cycles per day as a function of $\chi$, it is not sufficient to consider the approximation obtained by dividing the total number of sleep cycles over a large number of days by the total number of days. In the limit, this would give an accurate description of the average number of sleep cycles per day.  However, patterns in the number of sleep cycles per day can be of any length, and each model solution includes an initial region of transient behavior.  Thus, this approach is not practical computationally since very long simulations would be required to avoid errors introduced by early transient behavior. Instead, we need a method to detect a stable repeating pattern of the number of sleep cycles per day that does not include the initial period of transient behavior and can account for a variable length of the pattern. This approach has the additional advantage of establishing the pattern itself rather than simply determining the average number of sleep cycles per day. 
    
    For each pattern within a sequence, we specify the order of the pattern to be the length of the shortest subset of numbers that describes the pattern.  For example, the sequences $\{1,1,1,\ldots\}$ and $\{1,2,1,2, \ldots\}$ have order one and two, respectively, since they are generated by repeating patterns of $\{1\}$ and $\{1,2\}$, respectively.  Since our approach is computational, we cannot establish the existence of all possible patterns.  Instead, we focus on identifying the bifurcation structure in lower order patterns in order to infer the full bifurcation structure of the system. 
    
To find patterns in the number of sleep cycles per day, we proceeded as follows.  First we ran the model simulation over a 100 day period and determined the number of sleep cycles that started on each day. This generated a sequence of length 100 which reflected the number of sleep cycles per day over the simulation period.  We applied the algorithm to this sequence to determine the existence of a low order pattern within the sequence.  Since the system may undergo transient behavior before reaching a stable pattern in the number of sleep cycles per day, we required a stable pattern to exist over at least the last 25 days of the 100 day sequence rather than over the entire 100 day sequence.  Considering a 25 element sequence also reduces false detection of transient low order patterns.  For example, if a pattern such as $\{1,2,1,2,1,2,2\}$ was considered over only four days, the resulting sequence $\{1,2,1,2\}$ would mistakenly appear to have a stable repeating pattern of $\{1,2\}$.  Additionally, we require that the candidate pattern be repeated at least three times.  Since the algorithm only guarantees that sequence occurs during the last 25 days of the simulation and requires that it repeats at least three times, our ability to detect patterns of high order is restricted.  We let $n_{max}$ denote the longest length pattern that can be detected within our specified sequence length.  
    
    
    For $n=1, 2, \dots, n_{max}$, we searched the sequence for an $n$th order pattern as follows.  Let the first $n$ elements of the sequence to be the candidate pattern.  Determine if this pattern is repeated for the rest of the sequence. If so, this is designated to be the stable pattern. If not, apply the shift operator to the full sequence to provide the next candidate $n$th order pattern. Continue this process of updating the candidate pattern until a stable $n$th order pattern is identified or the number of elements remaining in the sequence becomes too small.  For higher order patterns, this may occur when the remaining sequence is shorter than 25 elements or if the total length of the pattern could not be repeated three times in the length of the remaining sequence.  If no stable pattern is detected before the candidate sequence becomes too long to check over the remaining sequence, then no stable $n$th order pattern is reported.  If no stable pattern is detected for $n \le n_{max}$, then the number of sleep cycles per day is approximated as the total number of sleep cycles in the simulation divided by 100 days.

\bibliographystyle{siamplain}
\bibliography{references}